# Phonon Transport in Single-Layer Transition Metal Dichalcogenides: a First-Principles Study

*Xiaokun Gu and Ronggui Yang\**

Department of Mechanical Engineering& Materials Science and Engineering Program,

University of Colorado at Boulder, Colorado 80309, USA

ABSTRACT:

Two-dimensional transition metal dichalcogenides (TMDCs) are finding promising electronic and optical applications due to their unique properties. In this letter, we systematically study the phonon transport and thermal conductivity of eight semiconducting single-layer TMDCs, $MX_2$ (M=Mo, W, Zr and Hf, X=S and Se), by using the first-principles-driven phonon Boltzmann transport equation approach. The validity of the single-mode relaxation time approximation to predict the thermal conductivity of TMDCs is assessed by comparing the results with the iterative solution of the phonon Boltzmann transport equation. We find that the phononic thermal conductivities of 2H-type TMDCs are above 50 W/mK at room temperature while the thermal conductivity values of the 1T-type TMDCs are much lower, when the size of the sample is 1 μm.

A very high thermal conductivity value of 142 W/mK was found in single-layer $WS_2$. The large atomic weight difference between W and S leads to a very large phonon bandgap which in turn forbids the scattering between acoustic and optical phonon modes and thus resulting in very long phonon relaxation time.

As a family of novel two-dimensional (2-D) materials beyond graphene, monolayer and a few layer transition metal dichalcogenides (TMDCs) have attracted considerable interests shortly after they were isolated or synthesized due to their unique physical properties and potential applications.[1-3] Generally monolayer TMDCs can have a three-layer structure that one layer of transition metal atoms are sandwiched by two layers of chalcogenide atoms. Depending on how the chalcogenide atoms are sitting on each side of the metal layer, there are two polymorphs for monolayer TMDCs: 1T phase with $D_{3d}$ point group and 2H phase with $D_{3h}$ point group, as shown in Fig. 1. The physical properties of molybdenum disulfide ($MoS_2$) with 2H structure, as a representative 2-D TMDC, have been widely studied. It exhibits a series of intriguing attributes different from its bulk form and from that of graphene, including the switchable thickness-dependent band gap,[4] strong photoluminescence[5] and significant anisotropic response under tensile strain.[6] In addition to $MoS_2$, other TMDCs with the same 2H crystal structure, such as $MoSe_2$, $WS_2$ and $WSe_2$, might be of similar or even superior properties to $MoS_2$. For example, these TMDCs are also of the thickness-dependent band gap.[7, 8] Triangular $WS_2$ monolayer displays strong room-temperature photoluminescence at the edge.[9] Compared with 2H TMDCs, 1T TMDCs gain relatively less notice, but might also possess some interesting properties. For example, 1T Zirconium and hafnium dichalcogenides are considered for photovoltaic application due to their suitable band gap for visible light absorption.[10, 11] In addition, both the bandgap of single-layer $ZrS_2$ can be effectively tuned by strain.[12]

Unlike the electronic, optical and mechanical properties of single-layer or few-layer TMDCs, which have been intensively explored, the study on the thermal properties has still been in its infancy, though its importance on the performance and reliability on the nano-devices are well recognized. According to the classical theory, the thermal conductivities of TMDCs are thought to be low due to their heavy atom mass and low Debye temperature.[13] This has led to the consideration of single-layer or few-layer TMDCs as potential thermoelectric materials.[14-16] While it is generally true that the cross-plane thermal conductivity of TMDCs are low due to the weak inter-layer bonding,[13] the in-plane thermal properties are unclear. There have recently been some studies on the in-plane thermal conductivity of monolayer or few-layer $MoS_2$.[17-25] While some classical molecular dynamics simulations using empirical interatomic potentials reported the thermal conductivity for the single-layer $MoS_2$ to be less than 10W/mK,[22, 23] the measured thermal conductivity values for the single-layer and multilayer $MoS_2$ are usually larger than 30 W/mK.[17-19] Considering the large uncertainty in the available thermal conductivity measurement data and the inaccuracy of the empirical potentials used in molecular dynamics simulations, the first-principles-based approach with predictive power has its unique strength to explore the phonon transport in 2-D TMDCs. A recent first-principles calculations under the single-mode relaxation time approximation (SMRTA) of the Peierls-Boltzmann transport equation (PBTE) suggested that the thermal conductivity of single-layer $MoS_2$ could even be higher than 70 W/mK when the size of sample is larger than 1 μm.[21] However, the validity of applying SMRTA to $MoS_2$ needs to be assessed since the phonon-phonon scattering is not elastic, though it is regarded as a reasonable approximation for some materials, such as silicon.[26]

In this letter, we present a systematic study of the phonon transport in single-layer TMDCs $MX_2$ (M = Mo, W, Zr and Hf, X = S and Se) by solving the PBTE with interatomic force constants inputs from first-principles calculations. The validity of applying SMRTA to predict the thermal conductivity of single-layer TMDCs is assessed first by comparing the calculation results of SMRTA and the iterative solution of the PBTE on $MoS_2$. The thermal conductivities of the eight single-layer TMDCs are then predicted from the iterative solution of the PBTE. Much higher thermal conductivity are found in 2H TMDCs, especially WS$_2$, comparing to that of 1T TMDCs. The origin of their distinct thermal transport properties are explored by detailed phonon scattering analysis.

In the first-principles-based approach, the accurate second-order harmonic and third-order anharmonic force constants are first extracted from density functional theory, which are employed to calculate the phonon transport properties, including phonon dispersion relation and three-phonon scattering rates (See Sec. S1, Supporting Information). Such phonon dynamics information is then used as the inputs for the PBTE, which considers the balance between phonon diffusion driven by the small temperature difference and phonon scatterings due to various scattering mechanisms. Here, we consider two kinds of phonon scattering mechanisms, three-phonon scattering and diffusive boundary scattering. The solution of PBTE provides the information of the population of each phonon mode and enables us to evaluate the thermal conductivity. The theoretical background of PBTE, including the phonon scattering mechanisms and the solution of PBTE from the SMRTA and the iterative approach, can be found in Sec. S2 of Supporting Information. Such first-principles-driven approach has been successfully used to predict the phononic thermal conductivity of a wide range of three-dimensional bulk crystals[27-31] and a few 2-D materials, such as graphene[32] and silicene.[33]

Table I summarizes the calculated lattice constants for the eight TDMCs studied in this work, which are in excellent agreement with the available measured monolayer[34] and the bulk lattice constants.[35-37] To report the values of thermal conductivity, the thicknesses of the monolayers, $h$, are also listed, which are defined as the measured cross-plane lattice constants or half of the lattice constants of the bulk materials, depending on 1T or 2H single-layer TMDCs. We also examine the bonding stiffness of these materials by calculating their spring constants as well as the sound velocities of the three lowest phonon branches, longitudinal acoustic (LA), transverse acoustic (TA) and flexural acoustic (ZA) branches, as listed in Table I. The spring constant, $K$, is defined as the trace of the harmonic force constant tensor of the nearest neighboring atom pairs (the metal atom M and the chalcogenide atom X), and written as[38] $K = \phi_{MX}^{xx} + \phi_{MX}^{yy} + \phi_{MX}^{zz}$, where $\phi_{MX}^{\alpha\alpha}$ is the second derivatives of the energy with respect to the displacement of atoms M and X along the Cartesian axis $\alpha$. Contradictory to the previous understanding that the bonding is weak in TMDCs, the bonding in single-layer molybdenum and tungsten dichalcogenides are indeed surprisingly stiff, even stiffer than silicon with a spring constant of 9.7 eV/Å. In general, the sulfides are 15% stiffer than the selenides, while the molybdenum dichalcogenides are 4% less stiff than tungsten dichalcogenides. Considering that the difference in the bonding strength in the group of 2H TMDCs is small, the mass of the basis atoms plays a key role in determine their phonon dispersion relations, which in term determines the related group velocity and thermal properties. Comparing to 2H TMDCs, the bonding in the 1T zirconium and hafnium dichalcogenides is about 50% weaker than their molybdenum and tungsten counterparts.

Figure 2 show the length-dependent thermal conductivity of MoS$_2$ calculated using both the iterative solution of PBTE and from SMRTA. Our SMRTA results are very close to Li *et al*'s calculations using a similar approach.[21] (The boundary scattering term in Ref. [21] is slightly different from our treatment. The compare the data directly, we scale the sample size from Ref. [21] by a factor of $\sqrt{2}$, as discussed in Sec. S2 of Supporting Information). However, the obtained thermal conductivity values from SMRTA are significantly smaller than that from strictly solving the PBTE iteratively. When the length of the monolayer sheet $L$ is smaller than 30 nm, the difference between two approaches is less than 5%. This is because the dominant phonon scattering comes from elastic boundary scattering when the concept of relaxation time is applicable. However, as the length increases where the phonon-phonon scattering becomes dominant, SMRTA cannot distinguish the resistive Umklapp process and the normal process, which does not directly provide the resistance to the heat flow. The under-prediction of SMRTA becomes distinguishable when the scattering due to normal process is strong. For example, when $L = 1$ μm, the thermal conductivity from the iterative solution of the PBTE is 103 W/mK, which is ~25% higher than the value of 83 W/mK from SMRTA. Due to the non-negligible difference between the SMRTA and the iterative solution for $MoS_2$, PBTE is strictly solved with the iterative approach in this work to accurately predict the thermal conductivity of $MoS_2$ and other TMDCs. Although SMRTA tends to underestimate considerably the thermal conductivity of single-layer TMDCs, the concept of phonon lifetime or scattering rate, that is used in SMRTA, of each phonon mode can still provide useful information on the strength of phonon-phonon scattering. We have thus still employ SMRTA when needed to qualitatively interpret the scattering mechanism in different materials.

The calculated thermal conductivity of $MoS_2$ using molecular dynamics (MD) simulations with empirical interatomic potentials[22, 23] is shown in Fig. 2. Clearly MD simulations have predicted a far too low thermal conductivity value comparing to the first-principles calculations. Although some of the potentials used in MD can reasonably reproduce the phonon dispersion, the anharmonicity was not taken into account when the empirical potentials were developed. The low thermal conductivity prediction from MD indicates that the anharmonicity in these empirical potentials has been overestimated

Figure 3 shows the calculated thermal conductivities of TMDCs with the sample size $L = 1 \mu m$ as a function of temperature, along with the available measurement data of single-layer TMDCs. The contribution from the three acoustic phonon branch and the optical branches are also calculated (See Supporting Information Sec. S3). Among the four single-layer 2H TMDCs, $WS_2$ is of the highest thermal conductivity, 142 W/mK at room temperature and then followed by $MoS_2$ (103 W/mK), $MoSe_2$ (54 W/mK) and $WSe_2$ (53 W/mK). It is notable that the thermal conductivity of $WS_2$ is the highest among all TMDCs studied and about 40% larger than that of $MoS_2$. The atomic mass of W is about twice as heavy as Mo. Table I shows that the bonding in $WS_2$ is only ~ 4% stiffer than that in $MoS_2$ according to the spring constants. The large thermal conductivity of $WS_2$ is contradictory to the classical theory which would expect a smaller phononic thermal conductivity due to the much heavier atom mass and weaker bonding stiffness.[39] Figure 4 shows the phonon dispersion of $MoS_2$ and $WS_2$. As expected, all the three acoustic branches of $WS_2$ are lower than that of $MoS_2$ due to the difference in atom mass and bonding stiffness between $MoS_2$ and $WS_2$. As a result, the group velocity and heat capacity of the acoustic phonons in $MoS_2$ are larger than $WS_2$, both of which facility the heat transport. However, much weaker phonon-phonon scattering is observed in $WS_2$ is than in $MoS_2$, especially for middle-range frequency phonon modes (50 $cm^{-1}$ to 200 $cm^{-1}$) by examining the phonon scattering rate $\Gamma$, as shown in Fig. 5.

In Figure 4, we also observe a very large frequency gap between the optical and acoustic phonon branches in $WS_2$, due to the large mass difference of the basis atoms of $WS_2$. The frequency gap is as large as 110 $cm^{-1}$, which is close to the range of acoustic phonons of $WS_2$ (178 $cm^{-1}$), while the gap is only 45 $cm^{-1}$ for $MoS_2$, much smaller than the range of acoustic phonons (230 $cm^{-1}$). Because of the large phonon frequency gap of $WS_2$, one important phonon scattering channel, the annihilation process of two acoustic phonon modes into one optical one (acoustic+acoustic->optical), becomes ineffective due to the requirement on energy conservation for phonon-phonon scattering, although such scatterings are not totally prohibited. The scattering through such scattering channel is usually the resistive Umklapp scattering.[40] As a result, the weaker phonon-phonon scattering rate is observed in $WS_2$ which renders to a much higher thermal conductivity.

To further show that the large phonon frequency gap leads to the large thermal conductivity of $WS_2$, we shift the phonon frequency of each optical phonon mode downward by the same amount to reduce the frequency gap, and then recalculate the thermal conductivity of the $WS_2$-like material. Figure 6 shows the calculated thermal conductivity as a function of the size of the phonon frequency gap. Clearly, the thermal conductivity monotonically decreases when the phonon frequency gap becomes smaller. In particular, when the frequency gap is the same as that of $MoS_2$, the thermal conductivity is reduced to 60 W/mK, a value even smaller than that in $MoS_2$. Recently, the first-principles calculations have been used to predict very high thermal conductivity of some three-dimensional bulk materials, such as $BAs$[40] and $AlSb$[30], primary due to a large frequency gap. Our simulations confirm that examining the acoustic-optical frequency gap could be a powerful search for 2-D materials with high thermal conductivity. We also plot results calculated from SMRTA in Fig. 6. The ratio between the thermal conductivities from the iterative solution and the SMRTA increases when the gap becomes large. This can be partially attributed to less resistive Umklapp scattering through the channel of acoustic+acoustic->optical. This observation confirms the importance of fully solving the PBTE to accurately predict the thermal conductivity of TMDCs.

Unlike the high thermal conductivity of 2H molybdenum and tungsten dichalcogenides, the thermal conductivities of zirconium and hafnium dichalcogenides are found to be much lower, ranging from 10 W/mK to 30 W/mK when the size of sample is 1 μm, as shown in Fig. 3(b). To explore the origin of the low thermal conductivity of these materials, we examine phonon dispersion and phonon lifetime of 1T TMDCs and compare them with 2H TMDCs. Figure 3(b) shows the phonon dispersion of $ZrS_2$ and $HfS_2$. The span of the phonon frequency is much smaller than 2H $MoS_2$ and $WS_2$, which could be attributed to the weak bonding stiffness. Figure 7 show the phonon lifetimes of both 2H and 1T TMDCs. While the phonon lifetimes of 2H TMDCs are all above 1 ps, the phonon lifetimes of 1T TMDCs are almost one order-of-magnitude smaller than that of 2H TMDCs. The strong scattering in 1T TDMCs is also correlated to their relatively small range of the phonon frequency. In 1T TMDCs, the separation between acoustic and the optical phonon branches is smaller, which results in much more frequent scattering between acoustic modes and optical modes. In addition, the strength of such scatterings is expected to be strong compared with the case in 2H TMDCs, because the population of the lower-frequency optical phonon modes involving the scattering with acoustic modes is larger according to the Bose-Einstein statistics, and the elements of the three-phonon scattering matrix is larger due to it inversely proportional relation with the phonon frequency.

In summary, we have used the first-principles-based PBTE approach to systematically predict the phononic thermal conductivity of eight typical single-layer TMDCs. The validity of the single-mode relaxation time approximation to predict the thermal conductivity of TMDCs is also assessed by comparing with the iterative solution of the phonon Boltzmann transport equation. We found that the thermal conductivities of $MoS_2$ and $WS_2$ are as high as 103 W/mK and 142 W/mK when the size of the sample is $1\mu m$, respectively. The large thermal conductivity of $WS_2$ can be attributed to the large acoustic-optical frequency gap due to the large mass difference of W and S, which makes inefficient scattering among acoustic and optical phonon modes. The thermal conductivities of 1T-type TMDCs are generally smaller than the 2H-type TMDCs due to the low bonding stiffness.

FIGURES

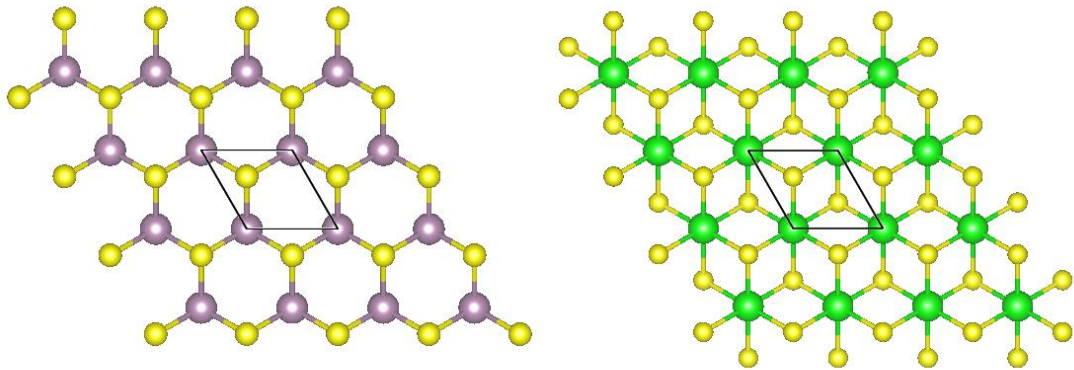

**Figure 1.** Crystal structures of (a) MoS$_2$ and (b) ZrS$_2$ monolayers as examples for 2H and 1T single-layer TMDCs, with the Mo atom in purple, the Zr atom in green, and the S atom in yellow.

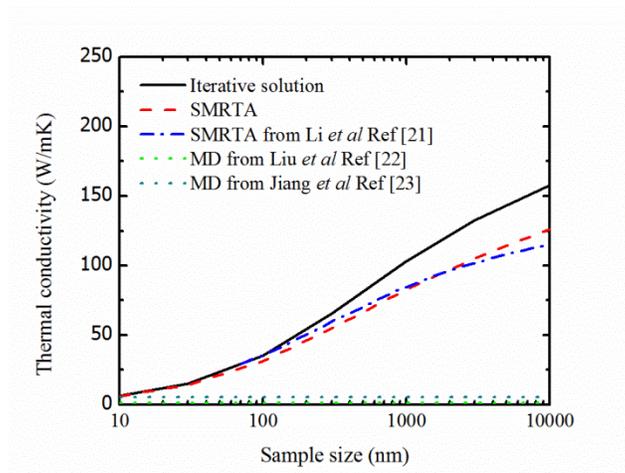

**Figure 2.** The calculated thermal conductivity of $MoS_2$ as a function of sample size.

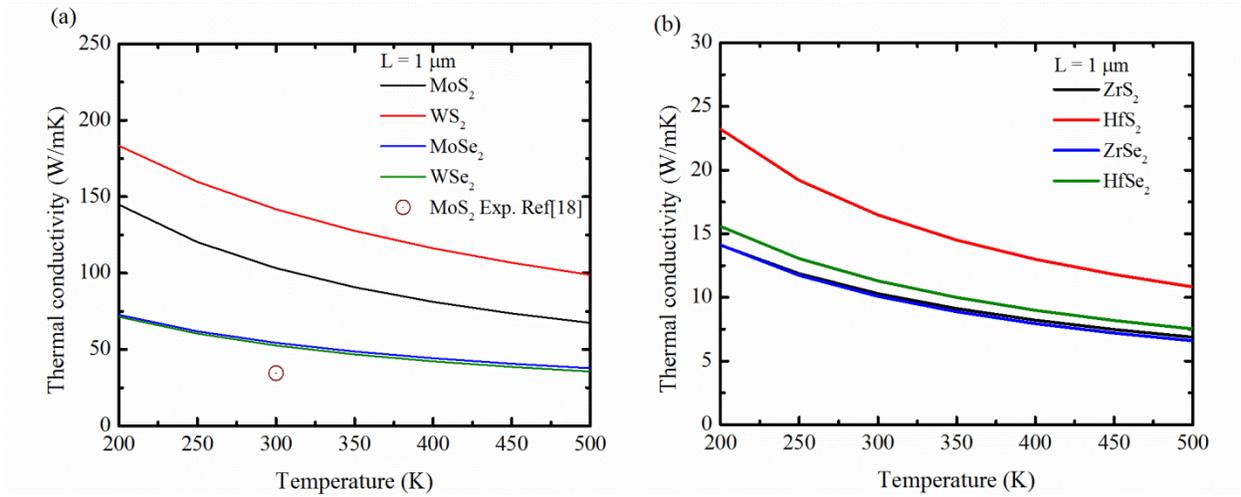

**Figure 3.** The thermal conductivity of (a) 2H and (b) 1T TMDC monolayers as a function of temperature.

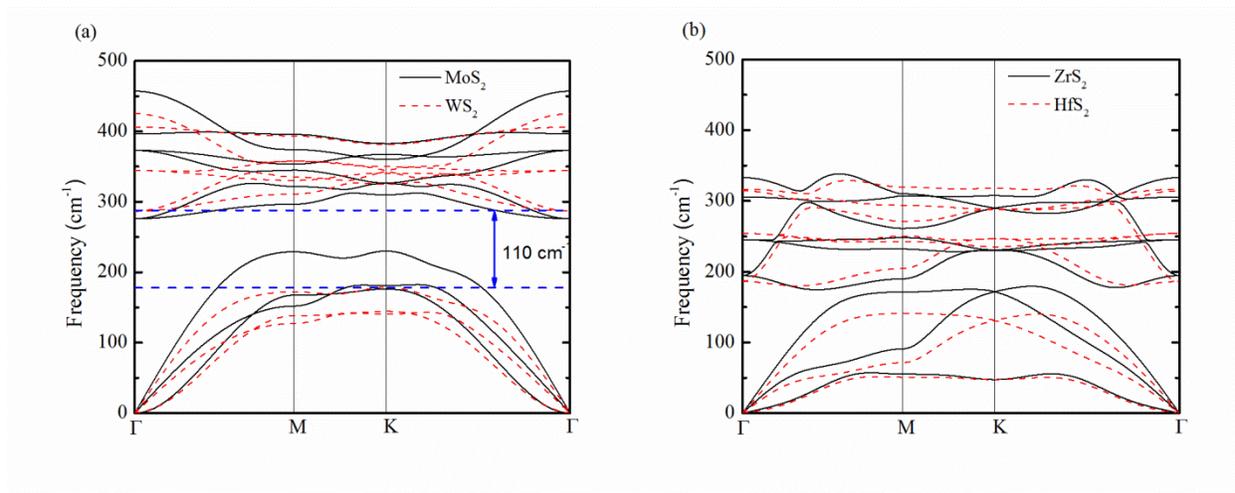

**Figure 4.** Phonon dispersion of (a) MoS$_2$ and WS$_2$, and (b) ZrS$_2$ and HfS$_2$ calculated from the first-principles simulations.

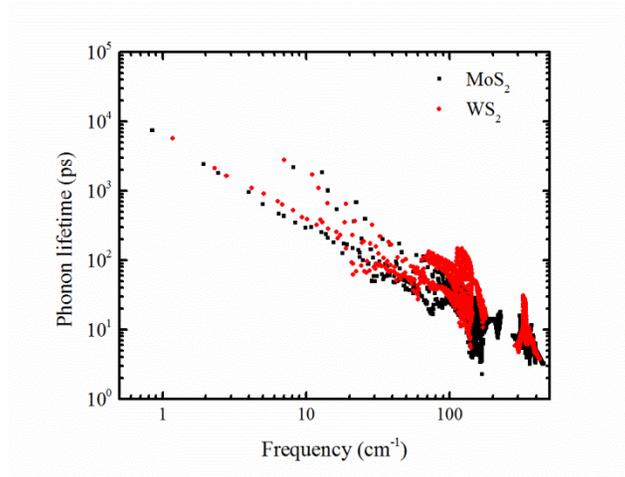

**Figure 5.** Phonon lifetime of $MoS_2$ and $WS_2$ at 300K as a function of phonon frequency.

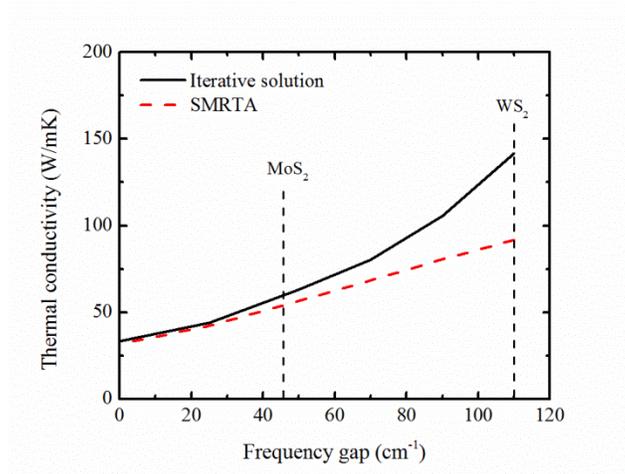

**Figure 6.** Calculated thermal conductivity of $WS_2$-like material at 300K as a function of the frequency gap between acoustic and optical branches. The black dashed lines indicate the frequency gap of $MoS_2$ and $WS_2$.

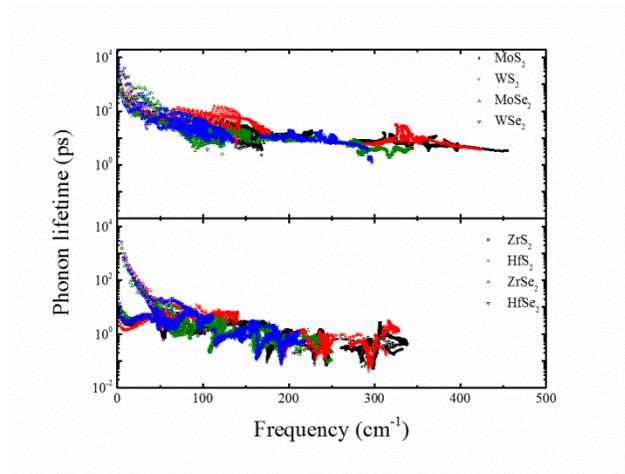

**Figure 7.** Phonon lifetime of (a) 2H and (b) 1T TMDCs at 300K as a function of phonon frequency.

TABLES.

**Table I.** Lattice constants, spring constants and sound velocity of single-layer TMDCs from DFT calculations, and the lattice constants from literature.

| material | lattice constant | | | thickness | spring constant | sound velocity | | |
|---|---|---|---|---|---|---|---|---|
| | $a$ (Å) (DFT) | $a$ (Å) (Exp. monolayer) | $a$ (Å) (Exp. bulk) | $h$ (Å) (Exp.) | $K$ (eV/Å) | $v_{LA}$ (m/s) | $v_{TA}$ (m/s) | $v_{ZA}$ (m/s) |
| $MoS_2$ | 3.19 | 3.22[a] | 3.16[b] | 6.15[b] | 11.2 | $6.4 \times 10^3$ | $4.1 \times 10^3$ | 480 |
| $WS_2$ | 3.19 | 3.23[a] | 3.15[c] | 6.16[c] | 11.7 | $5.5 \times 10^3$ | $3.5 \times 10^3$ | 610 |
| $MoSe_2$ | 3.32 | - | 3.30[b] | 6.47[b] | 9.8 | $4.8 \times 10^3$ | $3.0 \times 10^3$ | 340 |
| $WSe_2$ | 3.325 | 3.27[a] | 3.28[c] | 6.48[d] | 10.2 | $4.4 \times 10^3$ | $2.8 \times 10^3$ | 420 |
| $ZrS_2$ | 3.691 | - | 3.66[d] | 5.85[d] | 4.6 | $6.1 \times 10^3$ | $4.0 \times 10^3$ | 970 |
| $HfS_2$ | 3.646 | - | 3.62[d] | 5.88[d] | 5.2 | $5.0 \times 10^3$ | $3.3 \times 10^3$ | 630 |
| $ZrSe_2$ | 3.806 | - | 3.76[d] | 6.15[d] | 3.7 | $5.3 \times 10^3$ | $3.3 \times 10^3$ | 940 |
| $HfSe_2$ | 3.771 | - | 3.73[d] | 6.14[d] | 4.2 | $4.5 \times 10^3$ | $2.9 \times 10^3$ | 650 |

a Ref[34]

b Ref[35]

c Ref[36]

d Ref[37]

**Supporting Information**. Supporting Information about extracting interatomic force constants from first-principles calculations, the basics of Boltzmann transport equation and the contributions of the different phonon branches of 2H TMDCs are available. This material is available free of charge via the Internet at http://pubs.acs.org.

AUTHOR INFORMATION

**Corresponding Author**

*E-mail: ronggui.yang@colorado.edu

ACKNOWLEDGMENT

This work is supported by the NSF CAREER award (Grant No. 0846561), AFOSR Thermal Sciences Grant (FA9550-11-1-0109), and AFOSR STTR programs (PI: Dr. Sayan Naha). This work utilized the Janus supercomputer, which is supported by the National Science Foundation (award number CNS-0821794), the University of Colorado Boulder, the University of Colorado Denver, and the National Center for Atmospheric Research. The Janus supercomputer is operated by the University of Colorado Boulder. .

**Supplemental Information**

**for**

**Phonon Transport in Single-Layer Transition-Metal Dichalcogenides: a First-Principles Study**

Xiaokun Gu and Ronggui Yang*

Department of Mechanical Engineering

University of Colorado at Boulder, Colorado 80309, USA

*Address correspondence to: ronggui.yang@colorado.edu

# I. Extracting interatomic force constants from first-principles calculations

Our first-principles calculations are carried out with the Vienna *ab initio* Simulation Package (VASP)[1] with the projector augmented wave pseudopotential[2] with PBE functional. The kinetic-energy cut-off for the plane-wave basis set is set to be 500 eV and a $12\times12\times1$ k-mesh is used to sample the reciprocal space of the primitive unit cell. The choice of the energy cutoff and k-mesh ensures that the energy change is smaller than 1 meV/atom when refining these two parameters. To eliminate the interactions between periodic images of single layer samples in the first-principles calculations, a vacuum space of 2 nm is used. All materials are relaxed through the conjugate gradient algorithm until the atomic forces are smaller than $1\times10^{-5}$ eV/Å. The lattice constants, $a_0$, used in interatomic force constants and thermal conductivity calculations correspond to the lattice constants at zero-stress condition.

The standard direct method is employed to extract the harmonic and third-order anharmonic force constants from first-principles calculations.[3] We first record the forces of all atoms in a supercell when one or two atoms are displaced 0.015 Å away from their equilibrium positions, and then fit force-displacement data to extract both harmonic and anharmonic third-order force constants, using the following expression,

$$F_i^\alpha = -\sum_\beta \sum_j \phi_{ij}^{\alpha\beta} u_j^\beta - \frac{1}{2}\sum_{\beta\gamma}\sum_{jk} \psi_{ijk}^{\alpha\beta\gamma} u_j^\beta u_k^\gamma \tag{S1}$$

where $i,j,k$ and $\alpha,\beta,\gamma$ represent the index of atoms and the Cartesian coordinates, respectively; $u$ is the displacement of atom; $\phi$ and $\psi$ are the harmonic second-order and anharmonic third-order force constant. In crystal, the index can further been written as the pair $(R,\tau)$, with the position of the primitive unit cell **R** and the index of basis $\tau$ in the unit cell. The cutoffs of the harmonic and anharmonic interactions are chosen as $2.5\,a_0$ and $1.7\,a_0$, respectively. The cutoffs are tested on MoS$_2$ to ensure the converged phonon dispersion and thermal conductivity, as shown in Fig. S1 and S2. Two kinds of supercells with different dimensions ($4\times4\times1$ and $6\times6\times1$ primitive unit cells) are employed, where the numbers of k-points used are accordingly scaled down compared with the case of single-unit-cell calculation. While the calculations using larger-size supercells, where we only displace one atom, are essential to extract the long range harmonic force constants, the smaller-size supercells, where two atoms are displaced, are employed to extract the third-order anharmonic force constants with affordable computational resources. Although the smaller-size supercells are used, the long-range harmonic interactions are also taken into account in the fitting process by utilizing the boundary conditions.

With the harmonic force constants calculated from the first-principle calculations, the dynamical matrix $D$ with the pairs ($\tau,\alpha$) and ($\tau',\beta$) as indices is then solved for phonon dispersion,

$$D_{\tau\tau'}^{\alpha\beta}(\mathbf{q}) = \frac{1}{\sqrt{M_\tau M_{\tau'}}} \sum_{\mathbf{R'}} \phi_{0\tau,\mathbf{R'}\tau'}^{\alpha\beta} e^{i\mathbf{q}\cdot\mathbf{R'}} , \qquad (S2)$$

where $M_\tau$ is the atomic mass of the $\tau$th basis of the primitive cell. The phonon frequency $\omega_{\mathbf{q}s}$ is the square root of the $s$-th eigenvalue of the dynamical matrix and the group velocity $v_{\mathbf{q}s}^x$ is calculated as $\partial \omega_{\mathbf{q}s} / \partial q_x$.

Using the third-order force constants calculated from the first-principles, the three-phonon scattering rate can be calculated through the Fermi's golden rule. The three phonons involving the scattering have to satisfy the momentum conservation condition $\mathbf{q}\pm\mathbf{q}'=\mathbf{q}''+\mathbf{G}$, with $\mathbf{G}$ representing a reciprocal vector. When $\mathbf{G}=\mathbf{0}$ ($\mathbf{G}\neq\mathbf{0}$), the three-phonon process is the normal (Umklapp) scattering. The transition probabilities of the three-phonon processes $\mathbf{q}s+\mathbf{q}'s' \to \mathbf{q}''s''$ and $\mathbf{q}s \to \mathbf{q}'s'+\mathbf{q}''s''$ are written as[4]

$$W_{\mathbf{q}s,\mathbf{q}'s'}^{\mathbf{q}"s"} = 2\pi n_{\mathbf{q}s} n_{\mathbf{q}'s'} \left(n_{\mathbf{q}"s"}+1\right)\left|V_3\left(-\mathbf{q}s,-\mathbf{q}'s',\mathbf{q}"s"\right)\right|^2 \delta\left(\omega_{\mathbf{q}s}+\omega_{\mathbf{q}'s'}-\omega_{\mathbf{q}"s"}\right)$$

$$W_{\mathbf{q}s}^{\mathbf{q}'s',\mathbf{q}"s"} = 2\pi n_{\mathbf{q}s}\left(n_{\mathbf{q}'s'}+1\right)\left(n_{\mathbf{q}"s"}+1\right)\left|V_3\left(-\mathbf{q}s,\mathbf{q}'s',\mathbf{q}"s"\right)\right|^2 \delta\left(\omega_{\mathbf{q}s}-\omega_{\mathbf{q}'s'}-\omega_{\mathbf{q}"s"}\right), \quad (S3)$$

where the delta function denotes the energy conservation condition $\omega_{\mathbf{q}s} \pm \omega_{\mathbf{q}'s'} = \omega_{\mathbf{q}"s"}$ for the three-phonon scattering process, the + and - signs represent the annihilation and decay processes, respectively and $V_3$ is the three-phonon scattering matrix

$$V_3\left(\mathbf{q}s,\mathbf{q}'s',\mathbf{q}"s"\right) = \left(\frac{\hbar}{8N_0\omega_{\mathbf{q}s}\omega_{\mathbf{q}'s'}\omega_{\mathbf{q}"s"}}\right)^{1/2} \sum_{\tau}\sum_{\mathbf{R}'\tau'}\sum_{\mathbf{R}"\tau"}\sum_{\alpha\beta\gamma} \psi_{0\tau,\mathbf{R}'\tau',\mathbf{R}"\tau"}^{\alpha\beta\gamma} e^{i\mathbf{q}'\cdot\mathbf{R}'} e^{i\mathbf{q}"\cdot\mathbf{R}"} \frac{e_{\mathbf{q}s}^{\tau\alpha} e_{\mathbf{q}'s'}^{\tau'\beta} e_{\mathbf{q}"s"}^{\tau"\gamma}}{\sqrt{M_\tau M_{\tau'} M_{\tau"}}}.$$

(S4)

where $e$ is the eigenvector of the dynamical matrix, $N_0$ is the number of unit cells.

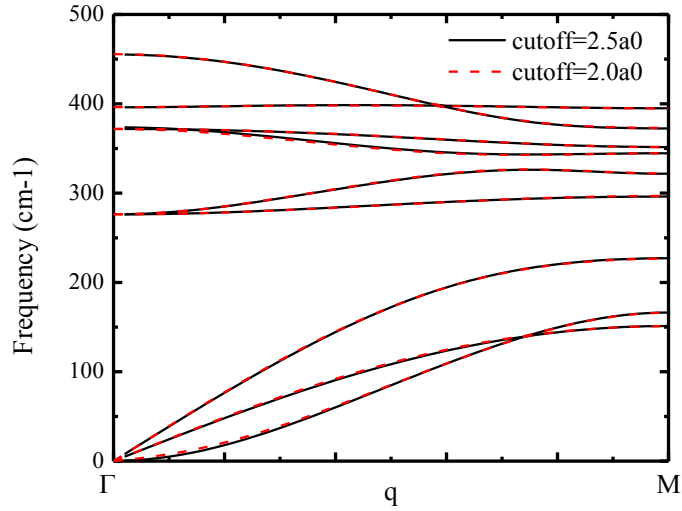

Figure S1. The calculated phonon dispersion of MoS$_2$ using different interaction cutoff for harmonic force constants.

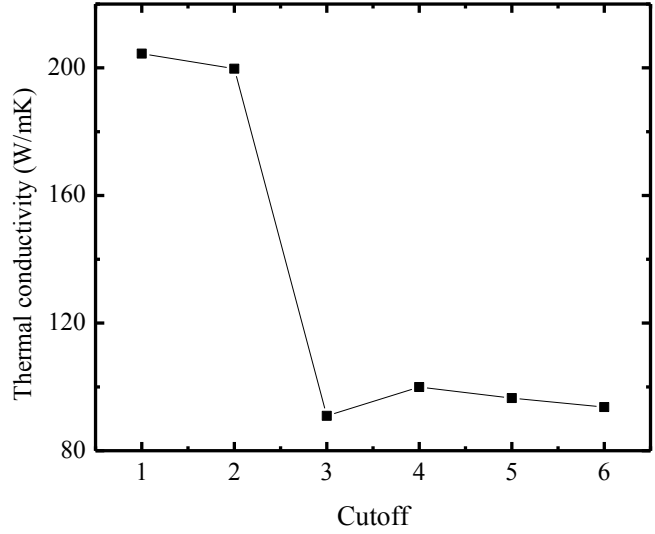

Figure S2. The calculated thermal conductivity of $MoS_2$ with the sample size of 1 μm as a function of the number of neighbor shells for third-order anharmonic interactions. The thermal conductivity is calculated using a $32\times32\times1$ sampling mesh in reciprocal space.

## II. Thermal conductivity calculation

Suppose the single-layer transition-metal dichalcogenides are lying in the *x-y* plane. A small temperature $\Delta T$ difference is applied to the two ends of the monolayer sheet with a distance $L$ apart in *x* direction. When the steady state is achieved, the heat flux can be expressed as the summation of the contributions from all phonon modes through

$$J = \frac{1}{(2\pi)^3} \sum_s \int \hbar \omega_{\mathbf{q}s} v^x_{\mathbf{q}s} n_{\mathbf{q}s} d\mathbf{q}, \tag{S5}$$

where $\mathbf{q}s$ stands for the *s*-th phonon mode at $\mathbf{q}$ in the first Brillouin zone, $\hbar$ is the Planck constant, and $n_{\mathbf{q}s}$ is the non-equilibrium phonon distribution function of mode $\mathbf{q}s$, respectively.

After *J* is calculated from the contributions of each phonon mode, the macroscopic thermal conductivity can then be calculated from the Fourier's law of heat conduction, $K_{xx} = J/(\Delta T/L)$.

## II.(a) Peierls Boltzmann transport equation (PBTE)

While the phonon modes obeys the Bose-Einstein (BE) distribution, $n_{\mathbf{q}s}^0$, in equilibrium condition, the phonon distribution function $n_{\mathbf{q}s}$ deviates from the BE distribution, which can be written as $n_{\mathbf{q}s}^0 + n_{\mathbf{q}s}^0 \left(n_{\mathbf{q}s}^0 + 1\right)\frac{\partial T}{\partial x} F_{\mathbf{q}s}$ with the unknown deviation function $F_{\mathbf{q}s}$,[5, 6] when temperature gradient is non-zero. The PBTE is used to solve the non-equilibrium phonon distribution function, or equivalently $F_{\mathbf{q}s}$. In this work, we consider two phonon scattering mechanisms, one is the boundary scattering, the other is the three-phonon scattering. The PBTE is expressed as

$$v_{\mathbf{q}s}^x \frac{\partial n_{\mathbf{q}s}^0}{\partial T} = \sum_{\mathbf{q}'s',\mathbf{q}''s''} \left[ \tilde{W}_{\mathbf{q}s,\mathbf{q}'s'}^{\mathbf{q}''s''} \left(F_{\mathbf{q}''s''} - F_{\mathbf{q}'s'} - F_{\mathbf{q}s}\right) + \frac{1}{2}\tilde{W}_{\mathbf{q}s}^{\mathbf{q}'s',\mathbf{q}''s''}\left(F_{\mathbf{q}''s''} + F_{\mathbf{q}'s'} - F_{\mathbf{q}s}\right) \right] - \frac{n_{\mathbf{q}s}^0\left(n_{\mathbf{q}s}^0+1\right)F_{\mathbf{q}s}}{L/2|v_{\mathbf{q}s}^x|}. \tag{S6}$$

In the expression, $\tilde{W}_{\mathbf{q}s,\mathbf{q}'s'}^{\mathbf{q}''s''}$ and $\tilde{W}_{\mathbf{q}s}^{\mathbf{q}'s',\mathbf{q}''s''}$ are the equilibrium transition probabilities for three-phonon annihilation and decay processes, respectively. They are the function of anharmonic third-order force constants, and the expressions can be found in ref. [7]. The last term in Eq. S(6) represents the boundary scattering. By comparing the wavelength of the dominant phonon modes from 200 K to 500 K and the roughness of the boundary, we expect that the boundary scattering is predominantly diffusive.[8] The relaxation time due to the fully diffusive boundary scattering is written as

$$\tau_{\mathbf{q}s}^{\mathrm{B}} = \frac{L}{2|v_{\mathbf{q}s}^x|}. \tag{S7}$$

In another empirical treatment of boundary scattering, the relaxation time due to boundary scattering could be written as

$$\tau_{\mathbf{q}s}^{B} = \frac{L'}{|v_{\mathbf{q}s}|}. \tag{S8}$$

with the group velocity $v_{\mathbf{q}s} = \sqrt{(v_{\mathbf{q}s}^x)^2 + (v_{\mathbf{q}s}^y)^2}$ and the characteristic length of the sample $L'$. The two treatments are not equivalent, but the thermal conductivities are almost identical if defining $L = \sqrt{2}L'$, according to the calculation on 2-D hexagonal BN.[9] The underlying mechanism of such coincidence is related to the isotropy of transport in 2-D hexagonal crystal.

## II.(b) Single-mode relaxation time approximation (SMRTA)

From PBTE, Eq. (S6), the population of each phonon mode is coupled with other phonon modes' population, which makes the PBTE difficult to solve. Under SMRTA, PBTE can be solved directly by assuming each phonon mode is decoupled with other modes, or $F_{\mathbf{q}'s'} = F_{\mathbf{q}''s''} = 0$. Then, the deviation function can be simply written as

$$F_{\mathbf{q}s} = -v_{\mathbf{q}s}^x \frac{\partial n_{\mathbf{q}s}^0}{\partial T} / \left\{ \frac{n_{\mathbf{q}s}^0(n_{\mathbf{q}s}^0+1)}{\tau_{\mathbf{q}s}^{ph}} + \frac{n_{\mathbf{q}s}^0(n_{\mathbf{q}s}^0+1)}{\tau_{\mathbf{q}s}^{B}} \right\}$$

$$\tau_{\mathbf{q}s}^{ph} = n_{\mathbf{q}s}^0(n_{\mathbf{q}s}^0+1) / \sum_{\mathbf{q}'s',\mathbf{q}''s''} \left( \tilde{W}_{\mathbf{q}s,\mathbf{q}'s'}^{\mathbf{q}''s''} + \frac{1}{2}\tilde{W}_{\mathbf{q}s}^{\mathbf{q}'s',\mathbf{q}''s''} \right) \tag{S9}$$

where $\tau_{\mathbf{q}s}^{ph}$ is the relaxation time due to three-phonon scattering and boundary scattering, respectively. Then, the thermal conductivity of under SMRTA is expressed as

$$K_{xx} = \frac{2\hbar^2}{N_0\sqrt{3}a_0^2 h k_B T^2} \sum_{\mathbf{q}s} \omega_{\mathbf{q}s}^2 (v_{\mathbf{q}s}^x)^2 n_{\mathbf{q}s}^0(n_{\mathbf{q}s}^0+1)\tau_{\mathbf{q}s}, \tag{S10}$$

with $\tau_{\mathbf{q}s} = 1/(1/\tau_{\mathbf{q}s}^{ph} + 1/\tau_{\mathbf{q}s}^{B})$, the thickness of TMDCs monolayer $h$, the Boltzmann constant $k_B$.

## II.(c) Iterative solution of PBTE

Apart from the SMRTA, the set of linear equations Eq. (S6), with respect to $F_{\mathbf{q}s}$, can then be self-consistently solved through iterative method. Here we employ the biconjugate gradient stabilized method (Bi-CGSTAB), a variant of the conjugate gradient algorithm,[10] to iteratively solve it. After $F_{\mathbf{q}s}$ is calculated, the thermal conductivity of the two-dimensional material can be written as

$$K_{xx}(x) = \frac{2\hbar}{N_0 \sqrt{3} a_0^2 h} \sum_{\mathbf{q}s} \omega_{\mathbf{q}s} v_{\mathbf{q}s}^x n_{\mathbf{q}s}^0 \left(n_{\mathbf{q}s}^0 + 1\right) F_{\mathbf{q}s} . \tag{S11}$$

By strictly solving Eq. (S6), the coupling between phonon modes are naturally taken into account. The thermal conductivity represented in the main text is calculated with dense meshes up to $80 \times 80 \times 1$ sampling points in reciprocal space.

## III. Thermal conductivity from different polarizations

To gain more insights on phonon transport in single-layer transition-metal dichalcogenides (TMDCs), we decompose the total thermal conductivity into the contributions of different phonon branches. Figure S3 shows the scaled thermal conductivity of LA, TA, ZA and optical branches for TMDCs as a function of temperature. The heat is almost evenly conducted by three acoustic branches, while the contribution of optical phonons is less than 5%, mainly due to less dispersive optical branches. In 2-D materials, the flexural acoustic phonon modes have attracted a lot of attention, partially inspired by their important role to the large thermal conductivity of graphene, where they conduct about 80% of the heat at room temperature.[7, 11] However, in TMDCs, only 30% of the heat is carried by the flexural phonon modes. The main difference between graphene and TMDCs is that graphene is one-atom-thick 2-D crystal while TMDCs are of sandwich-like structure. For the non-one-atom-thick 2-D crystals, such as TMDCs and silicene, a 2-D material with buckling structure, the symmetry selection rule in graphene that the flexural out-of-plane mode can only interacts with the other flexural out-of-plane mode and another in-plane mode, does not hold, as discussed in our previous paper.[7] Therefore, the phase space of three-phonon process for TMDCs' flexural mode is much larger than graphene, and thus the flexural modes are more likely to be scattered.

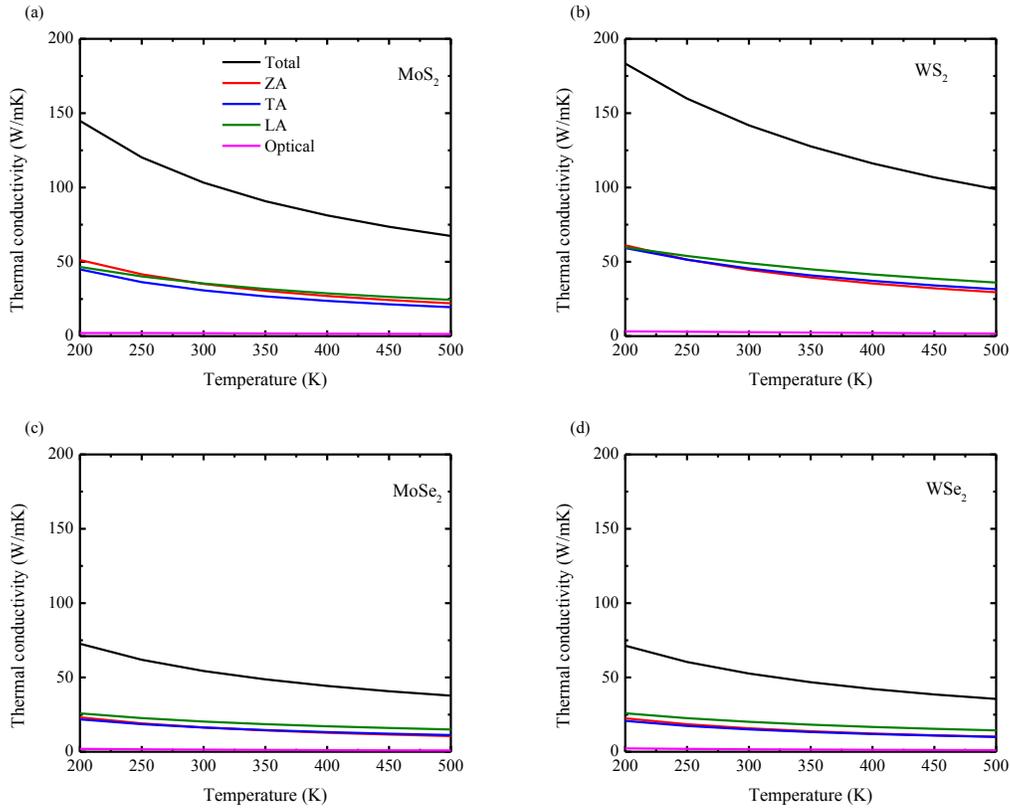

Figure S3. The thermal conductivity of 2H TMDCs from different phonon branches as a function of temperature.